%% file: paper.tex
\documentclass[
 aip,
 amsmath,amssymb,
 reprint,%
]{revtex4-1}

\usepackage{graphicx}
\usepackage{dcolumn}
\usepackage{bm}

\usepackage[utf8]{inputenc}
\usepackage[T1]{fontenc}
\usepackage{mathptmx}
\usepackage{etoolbox}
\usepackage{algorithm}
\usepackage{algpseudocode}
\usepackage{float}
\usepackage{subfig}
\usepackage{xcolor}

\makeatletter
\def\@email#1#2{%
 \endgroup
 \patchcmd{\titleblock@produce}
  {\frontmatter@RRAPformat}
  {\frontmatter@RRAPformat{\produce@RRAP{*#1\href{mailto:#2}{#2}}}\frontmatter@RRAPformat}
  {}{}
}%
\makeatother

\begin{document}

\title[Multifidelity Bayesian optimization for ICF]{A multifidelity Bayesian optimization\\ method for inertial confinement fusion design}
\author{J. Wang}
\author{N. Chiang}
\author{A. Gillette}
\author{J. L. Peterson}
\email{peterson76@llnl.gov}
\affiliation{Lawrence Livermore National Laboratory, Livermore, CA 94550}

\begin{abstract}
Due to their cost, experiments for inertial confinement fusion (ICF) heavily rely on numerical simulations to guide design.
As simulation technology progresses, so too can the fidelity of models used to plan for new experiments. However, these
high-fidelity models are by themselves insufficient for optimal experimental design, because their computational cost remains too high to efficiently and 
effectively explore the numerous parameters required to describe a typical experiment. Traditionally, ICF design 
has relied on low-fidelity modeling to initially identify potentially interesting design regions, which are then subsequently 
explored via selected high-fidelity modeling. In this paper, we demonstrate that this two-step approach can be insufficient: even for simple design problems, a two-step optimization strategy can lead high-fidelity searching towards incorrect regions and consequently waste computational resources on parameter regimes far away from the true optimal solution. We reveal that a primary cause of this behavior in ICF design problems is the presence of low-fidelity optima in distinct regions of the parameter space from high-fidelity optima. 
To address this issue, we propose an iterative multifidelity Bayesian optimization method based on Gaussian Process Regression that leverages both low- and high-fidelity modelings. 
We demonstrate, using both two- and eight-dimensional ICF test problems, that our algorithm can effectively utilize low-fidelity modeling for exploration, while automatically refining promising designs with high-fidelity models. This approach proves to be more efficient than relying solely on high-fidelity modeling for optimization.
\end{abstract}

\maketitle

\input{Sections_journal/Introduction.tex}

\input{Sections_journal/Bayesian.tex}

\input{Sections_journal/Examples.tex}

\input{Sections_journal/Conclusion.tex}

\appendix

\section*{Acknowledgments}
This work was performed under the auspices of the U.S. Department of Energy by Lawrence Livermore National Laboratory under contract DE--AC52--07NA27344 and the LLNL-LDRD Program under Project tracking No. 21-ERD-028.  Release number LLNL--JRNL--858267. The data that support the findings of this study are available from the corresponding author upon reasonable request.
\bibliography{bibliography}
\end{document}

%% file: Sections_journal/Introduction.tex
\newcommand{\norm}[1]{\left\lVert {#1} \right\rVert}
\newcommand{\Rbb}{\ensuremath{\mathbb{R} }}


\section{Introduction}\label{se:intro}

Design of experiments for inertial confinement fusion (ICF) is a challenging contemporary plasma science and engineering problem. And although the field has made great experimental strides in recent years, the problem of practical and robust ICF design has yet to be solved. Among the difficulties is the relative lack of experimental data~\cite{2022reviewplasma}. 
Leadership class experimental facilities are often limited to a few
experiments per week, such that even large campaigns may consist of only dozens of experiments. 
Therefore, numerical simulation has been a critical design tool for ICF: only numerically effective designs become candidates for testing in an actual experiment. 

Yet, full simulation-based design of ICF experiments remains a major challenge that requires the optimization of dozens of independent parameters, such as the timing and location of laser pulses, and the geometries and materials used in targets. And although the parameters are independent, the response surface is nonlinear, so that designers typically employ a select physics-informed manual or semi-automatic search in a design sub-space, instead of true optimal design in the entire space. 

Another challenge is that, even if employed in an optimization algorithm, each function evaluation is an expensive multiphysics simulation. A full indirect-drive coupled hohlraum-capsule simulation can cost a few node-days, and the simplest simulation that treats just a capsule with low-fidelity physics models can still take a few core-minutes~\cite{2022reviewplasma}.

Finally, since the design space is so large, any design found via local optimization (e.g. via gradient search using the simulator to estimate derivatives) is unlikely to be a robust solution: repeats of the exact same search are likely to get stuck in local minima and, therefore, find different solutions.

Recently, researchers have employed elaborate automated optimization workflows encompassing surrogate modeling, ensemble computing, machine learning, and gradient-free optimization algorithms to aid in numerical experimentation.
The ensemble computing approach of Peterson et al.~\cite{peterson2017} leveraged high-frequency ensemble computing and surrogate modeling to discover an optimized ICF design. 
However, the computational workflow to manage the ensembles was quite complex, requiring the automatic coordination, execution and post-processing 
of several thousand concurrently running independent HPC simulations. Using the Trinity supercomputer at Los Alamos National Laboratory to run several thousand lower-fidelity models, the researchers were able to train a random forest regression model, which was fast enough to embed into a global optimization algorithm. The new design parameters, suggested by optimizing the regression model, were then validated with new simulations.

An alternative to an ensemble approach is an iterative one, whereby simulations are run in small batches, with a model suggesting subsequent groups. Hatfield et al.~\cite{hatfield_genetic} used a genetic algorithm to evolve one-dimensional capsule simulations toward a new design. An iterative approach uses computational resources more efficiently, since new simulations are only placed in regions that are preferentially expected to improve the existing design.

Most recently, researchers have explored the use of multifidelity models for design optimization. The work of Vazirani et al~\cite{Vazirani:2023aa} blends an ensemble and iterative approach. First, they use low-fidelity (one-dimensional) capsule simulations to build a Gaussian Process regression algorithm that covers their entire design space. Then, they launch new higher-fidelity (two-dimensional) simulations and use the results of those simulations to train a multifidelity Gaussian Process (co-Kriging) model. This models serves as the driving engine for a Bayesian Optimization loop, which selects new two-dimensional simulations. The end result is an optimal two-dimensional design.

This two-step process allows for an initial exploration phase with low-fidelity models and then a subsequent refinement phase with high-fidelity models. Therefore, optima found in the first stage can guide the second stage, and high-fidelity models are spent only on regions that are likely to improve the high-fidelity solution.

Although this method saves computational resources by first exploring with low-fidelity models, it suffers from the same problem as ensemble approaches: in large parameter spaces it may be difficult (if not impossible) to build an adequate low-fidelity model that spans enough of the search space to meaningfully guide the second phase. Should the low-fidelity baseline be ill-representative of the true design surface (which is likely if under-sampled as one would expect in larger dimensional search spaces), the algorithm will only explore with subsequent high-fidelity simulations, which could become expensive. Furthermore, if the optimal locations of the low and high fidelity models are significantly different (e.g. by the physics differences between the models being sufficiently different so as to de-correlate their optima), the high-fidelity phase could waste time exploiting the wrong part of parameter space.

In this work, we demonstrate that for ICF problems, it is possible for the optima of the high and low-fidelity models to exist in different regions of parameter space. This decorrelation can bias high-fidelity models away from the true solution and limit the effective savings from a multifidelity approach.

To address this challenge, we propose a multifidelity Gaussian process Bayesian optimization method, novel for ICF design, that alternates between low and high fidelity models, combing data from both. Our algorithm allows for dynamic exploration by low and high-fidelity models, so that new low-fidelity simulations can influence (and therefore speed up) the high-fidelity search process. We demonstrate that this method is effective on both a two- and eight-dimensional design problem and that multifidelity Bayesian optimization is an effective tool for moderate-dimensional ICF design.

The paper is organized as follows. 
In Section~\ref{se:bayesian}, we describe the general Bayesian optimization method and our choice of kernel and acquisition function.
In Section~\ref{se:multi-bayesian}, we introduce the multifidelity Bayesian optimization model and propose our optimization algorithm.
Numerical experiments on two- and eight- dimensional ICF design test problems are shown in Section~\ref{sec:exp} that illustrate the capabilities of the proposed method.

%% file: Sections_journal/Bayesian.tex
\newcommand\ybm{{\ensuremath{\bm{y}}}}
\newcommand\xbm{{\ensuremath{\bm{x}}}}

\section{Bayesian Optimization}\label{se:bayesian}
\par\noindent
Consider the generic optimization problem: find $x^*\in \Rbb^n$ such that
\begin{equation} \label{eqn:opt-prob}
 \centering
  \begin{aligned}
   x^* ~=~&\underset{\substack{x\in C}}{\text{argmax}}
	  &  f(x),\\
  \end{aligned}
\end{equation}
where $C\subset \Rbb^n$ and $f:\Rbb^n\rightarrow \Rbb$.
Bayesian optimization is a computational technique for solving problem~\eqref{eqn:opt-prob} that does not require the objective function $f$ or its derivatives to have an analytical formula.
The method works in a sequential manner and treats $f$ as a ``black-box'' that can be queried at arbitrary inputs~\cite{frazier2018,bayesianoptreview2016}.
Bayesian optimization has been applied with great success in a wide variety of engineering settings, including inverse problems~\cite{wang2004}, structural design~\cite{mathern2021,wang2023optimization}, and robotics~\cite{calandra2016}. 
More recently, research has turned toward extensions of the method in the context of multifidelity surrogate models~\cite{zuluaga2013} and Gaussian processes with independent constraints~\cite{bernardo2011}.


 
In the context of ICF design, the function $f$ from (\ref{eqn:opt-prob}) is a map that takes a vector of design parameter values $x$ to a quantity of interest, e.g.,~nuclear yield.  
In a ``small'' design problem, $n$ may be $2$ or $3$, but a more informative design problem requires $7\leq n\leq 30$. 
Given a vector $x$, the value of $f(x)$ is determined by running an ICF simulation corresponding to $f$ determined by the parameter values indicated in $x$.

In Bayesian optimization, a subproblem is solved instead of~\eqref{eqn:opt-prob} at each iteration, where a surrogate function replaces $f$. 
A surrogate function is an approximation of $f$ constructed based on a collection of available simulation data. It is formed and updated at each iteration so that a more accurate approximation of $f$ can be used as the optimization progresses.   
In this paper, we use a Gaussian process method to create a multifidelity surrogate model of the objective function $f$, due to its flexibility of representing unknown functions and simplicity to implement.
At the input sample points $x$, the Gaussian process model assumes that there is a multivariate jointly Gaussian distribution among the input~$x$ and the design objective~$f$ as follows: 
\begin{equation} \label{eqn:GP-1}
 \centering
  \begin{aligned}
  \begin{bmatrix}
   f(x_1) \\
   \vdots\\
           f(x_T)
  \end{bmatrix} \sim \mathcal{N}\left( 
      \begin{bmatrix}    
                 m(x_1)\\
 \vdots\\
 m(x_T)
      \end{bmatrix},
              \begin{bmatrix}
               k(x_1,x_1) \dots k(x_1,x_T)\\
      \vdots\\
               k(x_T,x_1) \dots k(x_T,x_T)
      \end{bmatrix}\ 
      \right). 
\end{aligned}
\end{equation}
Here, $\mathcal{N}$ is the normal distribution with $T$ samples. The input design parameters or samples are denoted as $x_1,\ldots,x_T$. 
The function $m:\Rbb^N\to\Rbb$ is selected to provide a mean for the distribution and hence is called a ``mean'' function, even though it may not be an average. 
Finally, the function $k:\Rbb^n\times\Rbb^n\to\Rbb$ denotes the covariance function. 
Then, the posterior Gaussian probability distribution of a new sample point~$x$ can be inferred using Bayes' 
rule:
\begin{equation} \label{eqn:GP-post}
 \centering
  \begin{aligned}
  &f(x) | x, x_{1:T},f(x_{1:T}) \sim \mathcal{N} (\mu(x),\sigma^2(x))\\
  &\mu(x)\ =\ k(x,x_{1:T}) k(x_{1:T},x_{1:T})^{-1} \left(f(x_{1:T})-m(x_{1:T}) \right) + m(x_{1:T})\\
  &\sigma^2(x)\ =\
k(x,x)-k(x,x_{1:T})k(x_{1:T},x_{1:T})^{-1}\sigma(x_{1:T},x)\ ,
\end{aligned}
\end{equation}
where the vector~$x_{1:T}$ is the notation for $x_1,\dots,x_T$ and 
$k(x_{1:T},x_{1:T}) = [k(x_1),\dots,k(x_T); \dots; k(x_T,x_1),\dots,k(x_T,x_T)]$. 
The function $\mu$ and $\sigma^2$ are referred to as the posterior mean and variance, respectively.
It is common practice to set the mean function $m$ as a constant function; here we fix $m\equiv 1$.
The covariance functions (also known as the kernels) of the Gaussian process have significant impact on the accuracy of the surrogate model. 
A commonly used kernel, the Squared Exponential Covariance Function, also  
known as the power exponential kernel, is defined by 
\begin{equation} \label{eqn:kernel-sec}
 \centering
  \begin{aligned}
  k(x,x';\theta)\ =\
\text{exp}\left(-\frac{\norm{x-x'}^2}{\theta^2}\right)\ ,
  \end{aligned}
\end{equation}
where~$\theta$ denotes the hyper-parameters of the kernel. The hyper-parameters of the Gaussian process surrogate model 
 are often optimized during the fitting step by maximizing the 
log-marginal-likelihood with a gradient-based optimization method, such as the popular 
 L-BFGS algorithm~\cite{optimization}. 

Selection of a suitable acquisition function is also critical to an efficient and effective Bayesian optimization algorithm
and should reflect the user's preference in the trade-off between exploration and exploitation. 
If the acquisition function lays more emphasis on exploration, then maximizing it tends to return points with high uncertainty. Otherwise, one can expect a higher objective function value with a higher acquisition function value. 
In the case of a Gaussian process surrogate model, exploration and exploitation correspond to variance and mean values, respectively. 
One popular acquisition function is ``expected improvement" (EI)~\cite{brochu2010}, whose form can be written as
\begin{equation} \label{eqn:EI-1}
 \centering
  \begin{aligned}
       EI(x)\ =\ \begin{cases}
       (\mu(x) - f(x^+) - \xi)\Phi(Z)+\sigma(x)\phi(Z), &\text{if} \ \sigma(x)>0\\
       0, &\text{if} \ \sigma(x)\ =\ 0    
                 \end{cases}
  \end{aligned}
\end{equation}
where 
\begin{equation} \label{eqn:EI-2}
 \centering
  \begin{aligned}
       Z\ =\ \begin{cases}
       \frac{\mu(x)-f(x^+)-\xi}{\sigma(x)}, &\text{if} \ \sigma(x)>0\\
       0, &\text{if} \ \sigma(x)\ =\ 0\ .    
                 \end{cases}
  \end{aligned}
\end{equation}
The variable~$x^+$ in~\eqref{eqn:EI-1} is defined as 
$x^+=\text{argmax}_{x_i\in x_{1:T}}f(x_i)$ of all the existing $T$ 
samples. The trade-off parameter~$\xi\geq 0$ controls the trade-off between global 
search and local optimization~\cite{lizotte2008}. At an input sample point~$x$, the mean value predicted by 
the surrogate is~$\mu(x)$ and the variance is~$\sigma(x)$. The functions 
$\phi$ and~$\Phi$ are the probability density function (PDF) and cumulative 
distribution function (CDF) of the standard normal distribution, respectively. 
Expected improvement can be extended to constrained optimization problems, by 
introducing penalties to present the constraints in the objective~\cite{gardner2014}.
Thanks to its well-tested effectiveness, we select expected improvement as our acquisition function for the high fidelity model. 

The next sample point in the Bayesian optimization algorithm is chosen by maximizing the acquisition function over $C$. 
At the $k$th iteration, this means 
\begin{equation} \label{eqn:acquisition-1}
 \centering
  \begin{aligned}
      x_{k} = \text{argmax}_{x\in C}  EI_k (x),
  \end{aligned}
\end{equation}
where $EI_k$ is the expected improvement function using samples accumulated up to the $k$th iteration.
In order to solve~\eqref{eqn:acquisition-1}, optimization algorithms including L-BFGS or random 
search can be used~\cite{}.

The Bayesian optimization algorithm is given in Algorithm~\ref{alg:bo}.
\begin{algorithm}[H]
 \caption{Gaussian process Bayesian optimization for ICF design}\label{alg:bo}
  \begin{algorithmic}[1]
	  \State{Choose initial sampling points $x_i,~i\in[0,N_0]$.} 
	  \State{Train a GP surrogate model on the initial data}
  \For{$k=0,1,2,\dots$}
	  \State{Evaluate the acquisition function and find the new sample point $x_{k} = \text{argmax}_{x \in C} EI_k(x)$.}
	  \State{Run multiphysics ICF simulation with new design parameters $x_k$.}
	  \State{Compute the objective $f(x_{k})$ based on the simulation result. \;}
	  \State{Retrain the GP surrogate model with the addition of the new sample $x_k$ and $f(x_k)$.\;}
	  \State{Solve the optimization problem~\eqref{eqn:opt-prob} with the updated surrogate model.
	  \If {Stopping criteria satisfied} Exit
	  } 
  \EndFor
  \end{algorithmic}
\end{algorithm}
Line $7$ retraining the surrogate model refers to finding the hyperparameters of the surrogate model. For simplicity, line $8$ of Algorithm~\ref{alg:bo} can be solved through exhaustive search on a fixed, sufficiently large set of sample points in $C$. If the new maximum is found near the same point as the previous iteration or the difference in maximum value is smaller than a user-defined error tolerance, then the algorithm exits. Alternatively, upon reaching a prescribed maximum number of iterations, the algorithm exits. 

Algorithm~\ref{alg:bo} is implemented in python using the \texttt{scikit-learn} machine learning library~\cite{scikit-learn,sklearn_api} and uses `\texttt{standardscalar}' 
preprocessing to scale the objective in all fidelities to a similar range. We note that the Bayesian optimization formulation is presented as a maximization algorithm. 
Therefore, we take the negative value of the objective $f$ in~\eqref{eqn:opt-prob} when applying gradient-based optimization methods such as BFGS. 
%
%

\section{Multifidelity Bayesian optimization}\label{se:multi-bayesian}
Our multifidelity Bayesian optimization method uses surrogate models 
that rely on simulation results from more than one fidelity. In ICF design problems, where low-fidelity simulations are 
much less computationally expensive to perform than higher-fidelity simulations, a multifidelity approach that explores with cheaper (and less accurate) simulations and only selectively chooses more expensive simulations could lead to significant efficiency gains. Because the low-fidelity simulations should influence the high-fidelity search, one needs a multifidelity surrogate model that considers information from all fidelities. In this section, we extend the Bayesian optimization algorithm to such a multifidelity approach.

Let $N$ be the number of fidelities, and superscript $0$ denotes the index for the lowest fidelity. In this work, the $i$th fidelity surrogate model is chosen as 
\begin{equation}
  f^0(x) ~\text{and}~f^i(x) = \rho^{i-1}f^{i-1}(x) + \delta^{i-1},~1 \leq i < N,
\end{equation}
where $\rho^{i-1}$ is a scalar and $\delta^{i-1}$ is itself a Gaussian process.
The lowest-fidelity surrogate model $f^0$ is a Gaussian process, which is reasonable under a Markov assumption~\cite{kennedy2000}.
In the case of $N=2$, one can obtain the closed form expression for the predictive distribution through the co-kriging process and a prior assumption, \textit{e.g.}, non-informative `Jeffreys priors'. For the purpose of clarification, when $N=2$, we use superscript $l$ and $h$ to denote the low fidelity and high fidelity, respectively.

With $N=2$, the surrogate model $f^h$ is used to predict the high-fidelity model (and therefore the optimization objective $f$), and its accuracy can be verified with high-fidelity simulation results. The parameters that specify the surrogate models are the mean values $\mu^l,\mu^h$, variances $(\sigma^l)^2,(\sigma^h)^2$ of each Gaussian process, the scalar $\rho$ and hyperparameters $\theta^l,\theta^h$ in the kernel functions.
All of these model parameters need to be computed and updated at each optimization iteration.
Instead of a fully Bayesian approach, which takes into account the uncertainty of the model parameters themselves, 
we opt for a Bayesian estimation of the parameters $\mu^l,\mu^h$, $(\sigma^l)^2,(\sigma^h)^2$ and $\rho$ for efficiency~\cite{legratiet2013}. 
At each level, the kernel hyperparameter $\theta^l,\theta^h$ is obtained through minimizing the negative concentrated restricted log-likelihood with gradient-based optimization methods such as L-BFGS, as is mentioned in the single fidelity case in Section~\ref{se:bayesian}. 

As the dimension of the design space grows, the difficulty of finding optimal hyperparameters $\theta^l$ and $\theta^h$ increases as well, since we do not have access to analytic gradients, and it becomes increasingly possible for the L-BFGS algorithm to become stuck in local minima.  
To address this, we applied L-BFGS with multiple random starting points in hopes of finding the global minimizer. 
For more details on computing the closed form posterior distribution, we refer readers to Ref.~[\onlinecite{legratiet2013}].

For the high-fidelity surrogate model, we select new points by optimizing the expected improvement acquisition function~\eqref{eqn:EI-1}. 
Meanwhile, the low-fidelity sampling points are chosen based on a mutual-information acquisition function, as suggested in Refs.~[\onlinecite{contal2014}] and [\onlinecite{sarkar2019}].
Compared to expected improvement, this function is more likely to escape a local minimum and leans mathematically towards 
exploration during early stages of the optimization~\cite{contal2014}. As the surrogate model becomes more accurate in its prediction, exploitation is emphasized more. In this fashion, low-fidelity simulations will explore early on and then exploit later, which allows them to refine the high-fidelity solution. Since expected improvement often greedily exploits, the combined algorithm sets up a system that preferentially uses low-fidelity simulations for exploration and high-fidelity simulations for exploitation.

Suppose the low-fidelity Gaussian process surrogate model predicts at iteration $k$ the mean value $\mu_{k}^l$ and variance $(\sigma_{k}^l)^2$,
the exploration part of mutual information is governed by 
  \begin{equation}\label{eqn:GPMI-1}
    \begin{aligned}
	    \phi_k(x) = \sqrt{\alpha}\left(\sqrt{(\sigma^l_k)^2(x) +\hat{\gamma}_{k-1}} -\sqrt{\hat{\gamma}_{k-1}} \right).
    \end{aligned}
  \end{equation}
  The new low-fidelity sampling point is chosen to be  
  \begin{equation}\label{eqn:GPMI-2}
    \begin{aligned}
	    x_k^l  = \text{argmax}_{x\in C} \mu_{k}^l(x)+\phi_k(x).
    \end{aligned}
  \end{equation}
The quantity $\hat{\gamma}_k$ forms a lower bound on the information acquired at iteration $k$ and is updated by  
  \begin{equation}\label{eqn:GPMI-3}
    \begin{aligned}
	    \hat{\gamma}_k = \hat{\gamma}_{k-1} + (\sigma_k^l)^2(x_k^l)
    \end{aligned}
  \end{equation}
The trade-off between exploration and exploitation is controlled by the user-defined parameter $\alpha > 0$. 

The multifidelity Bayesian optimization algorithm with two fidelity levels is given in Algorithm~\ref{alg:mfbo}. At all iterations $k$, we maintain $x_k^h\subset x_k^l$, i.e., all the high-fidelity samples are also included in low-fidelity samples. Moreover, for each high-fidelity simulation performed, a sequence of additional low-fidelity simulations may be performed during the same iteration. 

A positive user-defined integer parameter $N_\ell\geq 1$ is used to denote the ratio between the low-fidelity sample points added into the training of the model and the high-fidelity sample point added at each iteration $k$. This tunable ratio is important for performance, because it lets the user sync the completion of the high and low fidelity simulations of each iteration, base on the relative cost of each. For instance, if the high-fidelity models costs 10 times the computational resources as the low fidelity model, setting $N_\ell=10$ would allow the low-fidelity simulations to finish at roughly the same time as the high-fidelity simulation. (Although of course the exact wallclock timing depends on the available parallel computational resources, which further underscores the need for a user-defined parameter.)

When $N_\ell = 1$, one high-fidelity and one low-fidelity sample point are being added at iteration $k$, though at the same input $x$, which is determined by maximizing expected improvement. Furthermore, the algorithm skips the low-fidelity loop from line 8 to line 13.
\begin{algorithm}[H]
 \caption{Multifidelity Bayesian optimization with Gaussian process for ICF design}\label{alg:mfbo}
  \begin{algorithmic}[1]
	  \State{Choose initial sampling data for low fidelity model $x_{0}^l$ and high fidelity model $x^h_{0}$.} 
	  \State{Build initial multifidelity surrogate model}
  \For{$k=1,2,\dots$}
	  \State{Evaluate acquisition function and find the new sample point $x_k^h = \text{argmax}_{x\in C} EI_k(x)$.}
	  \State{Run ICF simulations of both fidelities with $x_k^h$.}
          \State{Compute $f^h(x_k^h)$ and $f^l(x_k^h)$ based on the simulation results.\;}
	  \State{Retrain the surrogate model, including $x_k^h$ and $f^h(x^h_k), f^l(x^h_k)$.\;}
        \For{$j=1,2,\dots,N_\ell-1$}
	    \State{Evaluate acquisition function and find $x_j^l$ through~\eqref{eqn:GPMI-2}.  } 
	  \State{Run ICF simulation of low fidelity with $x_j^l$.}
          \State{Compute $f^l(x_j^l)$ based on the simulation result. \;}
	  \State{Retrain the surrogate model, including $x_j^l$ and $f^l(x_j^l)$.\;}
	\EndFor
	  \State{Solve the optimization problem~\eqref{eqn:opt-prob} with the updated surrogate model.
	  \If{Stopping criteria satisfied} Exit.} 
  \EndFor
  \end{algorithmic}
\end{algorithm}
The optimization step above (line 14) is executed similarly to that of Algorithm~\ref{alg:bo} (line 8) and the stopping criteria remains the same. The retraining steps on line 7 and 12 refer to finding the optimal hyperparameters of the corresponding surrogate models.

%% file: Sections_journal/Examples.tex
\section{Numerical Experiments}\label{sec:exp}

We have implemented Algorithms~\ref{alg:bo} and~\ref{alg:mfbo} in Python using scikit-learn, PyTorch and other open source code from OpenMDAO~\cite{Gray2019a}. 
We now describe our numerical experiments in two ICF design optimization model problems, using either two or eight design parameters.
In both cases, the objective function is based on simulation data collected using HYDRA, a multi-physics simulation code developed at LLNL~\cite{Metal2001}.
To perform a statistical study of the efficiency of the algorithm, we don't test our algorithm on HYDRA itself, but rather on pre-trained neural networks fitted to a HYDRA simulation database of one-dimensional capsule simulations of design variations around NIF shot N210808. This allows us to run the same optimization experiment several times to gather data on the statistical performance of the algorithm (a process that would be too expensive if running HYDRA natively).

The high-fidelity database consists of standard simulations but the low-fidelity simulations are considered ``burn off," in which the nuclear cross section has been artificially reduced by a factor of 1000, effectively turning off any yield amplification from alpha particle deposition. This ``burn-off" model, while adopted as a simple simulation tool to create a low-fidelity physics model, actually has real world applications: artificially dudded THD implosions (which have hydrogen added to the DT fuel) have been proposed as a means to speed up NIF operations and experimental studies for high-yield designs. In theory, if one can find the best design with a dudded implosion and then shoot that best design with a normal fuel layer, an experimental campaign could be sped up, since low-yield operations can be fielded at a faster cadence than high-yield operations.

\subsection{2D Design Example}

In our first example, we select design variables `\texttt{t\_2nd}' and `\texttt{sc\_peak}', which are the timing of the second shock and the strength of the peak of the radiation drive. The design objective $f$ is the nuclear yield. 
It is possible to verify and visualize this 2D example, and we set a fixed sample ratio $N_{l}=2$.
For the remaining parameters, we set $\xi=0$ (from (\ref{eqn:EI-1})) and $\alpha=10$ (from (\ref{eqn:GPMI-1})).

\begin{figure}
	\subfloat[Low fidelity ground truth.]{\includegraphics[width=0.48\columnwidth]{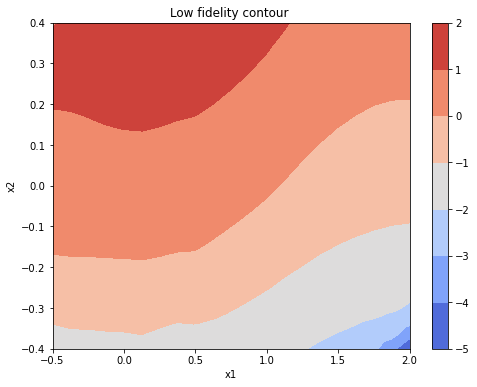}}
	\subfloat[High fidelity ground truth.]{\includegraphics[width=0.48\columnwidth]{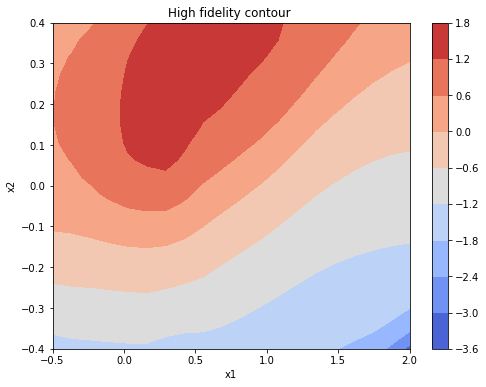}}\\
	\subfloat[Contour plots at iteration 14. Black line: location of new high and low-fidelity points.]
{\includegraphics[width=0.96\columnwidth]{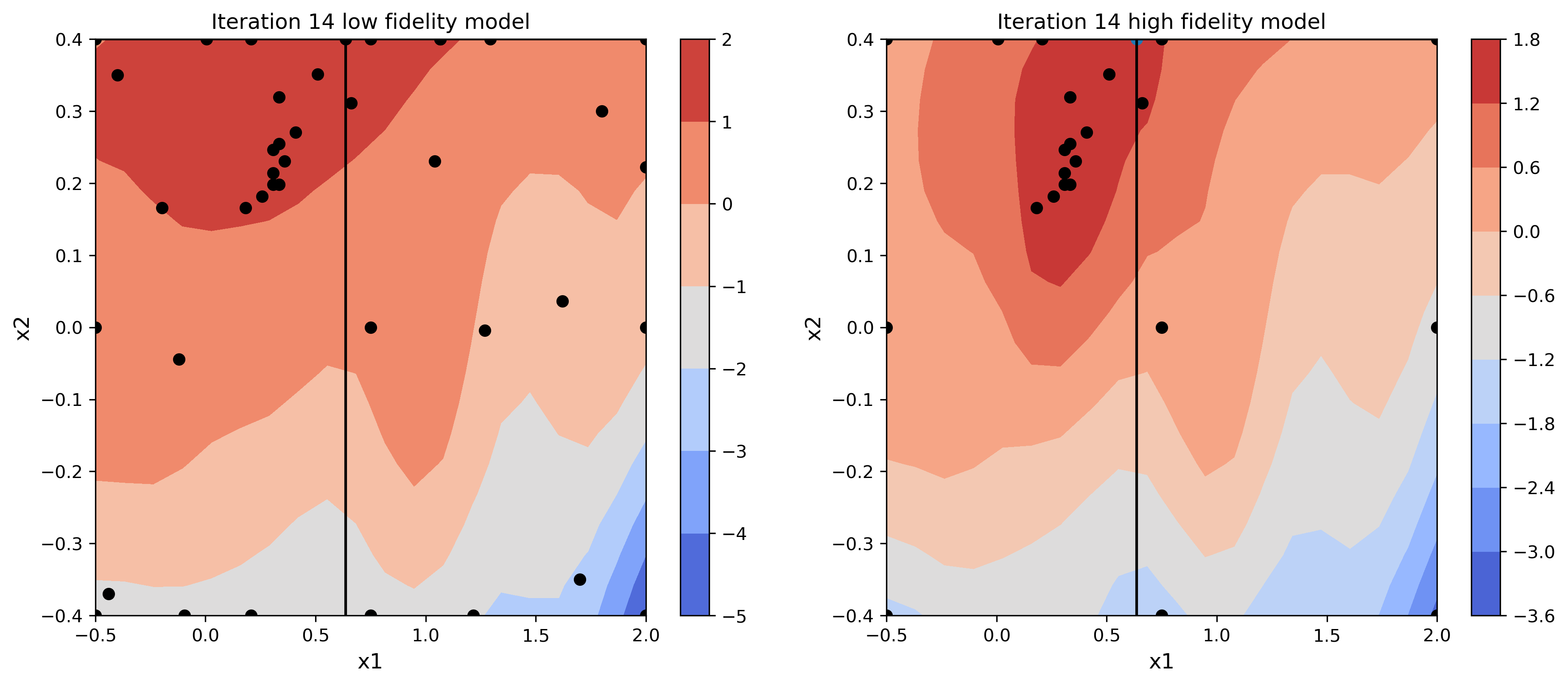}}\\
	\subfloat[Contour plots at iteration 14, low-fidelity loop. Black line: new low-fidelity point.]
{\includegraphics[width=0.96\columnwidth]{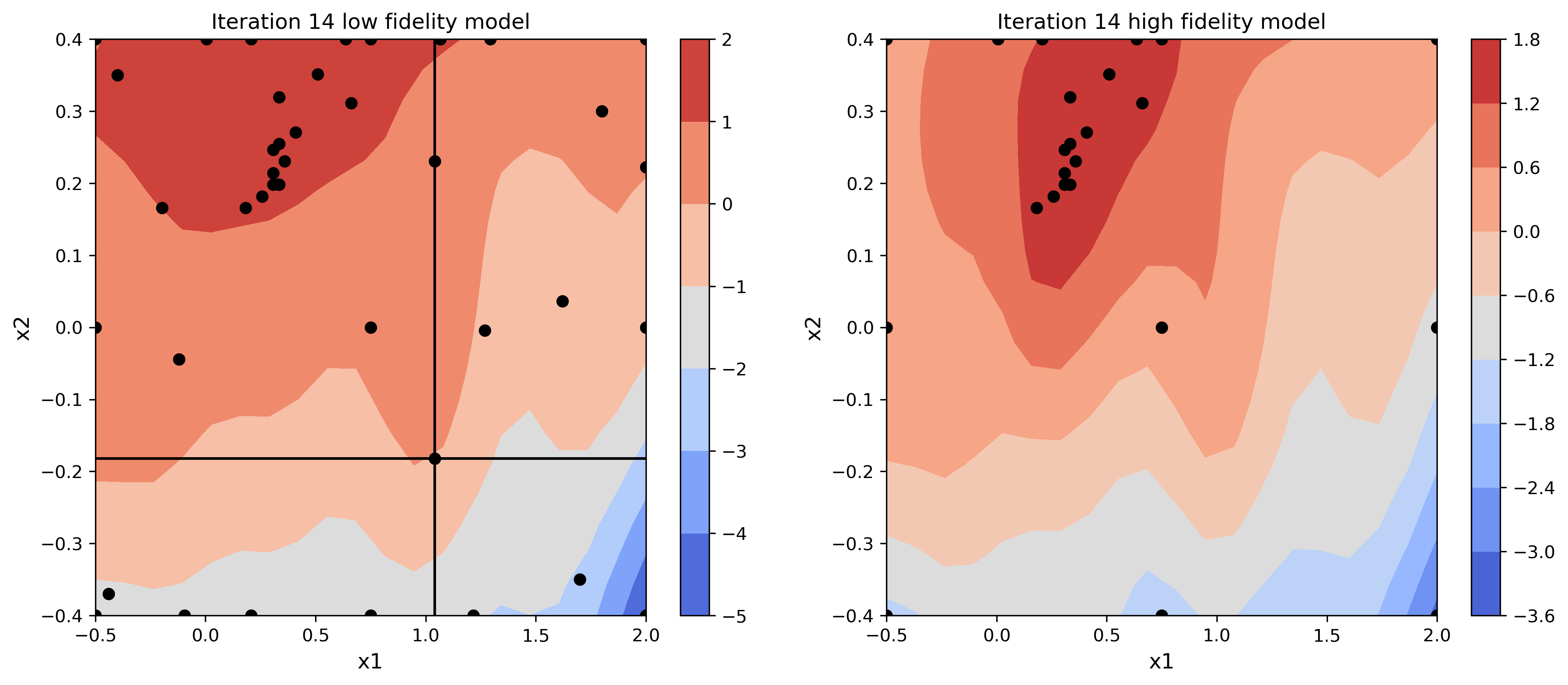}}\\
  \caption{Contour plots for the 2D design example, with existing simulations as black dots. The ground truth solution (a and b) show that the optimal points for low and high fidelity are in different parts of parameter space. The algorithm alternates between selecting a pair of high and low fidelity points (c) and just additional low-fidelity points (d).}
  \label{fig:ex1}
\end{figure}

Exhaustive sampling is possible in two-dimensions, which we use to determine ground truth.
In Figure~\ref{fig:ex1}, we show contour plots of the scaled output nuclear yield over the 2D design space.
It is clear from the ground truth plots (a) and (b) that the low- and high-fidelity simulation models predict both different maximum outputs and distinct optimal design variables, making the multifidelity approach model non-trivial. 

We apply our multifidelity Bayesian optimization algorithm to this problem, where the bound constraint $C$ is set to the upper and lower bounds for each design parameter. The algorithm successfully terminates in 15 iterations. The final mean value function contour and all the sampling points throughout the iterations for both fidelities are shown in Figure~\ref{fig:ex1} (c) and (d). 
Compared to Figure~\ref{fig:ex1}, the optimal objective and design variables predicted by the learnt surrogate model are accurate.

\begin{figure}
  \centering
	\includegraphics[width=0.96\columnwidth]{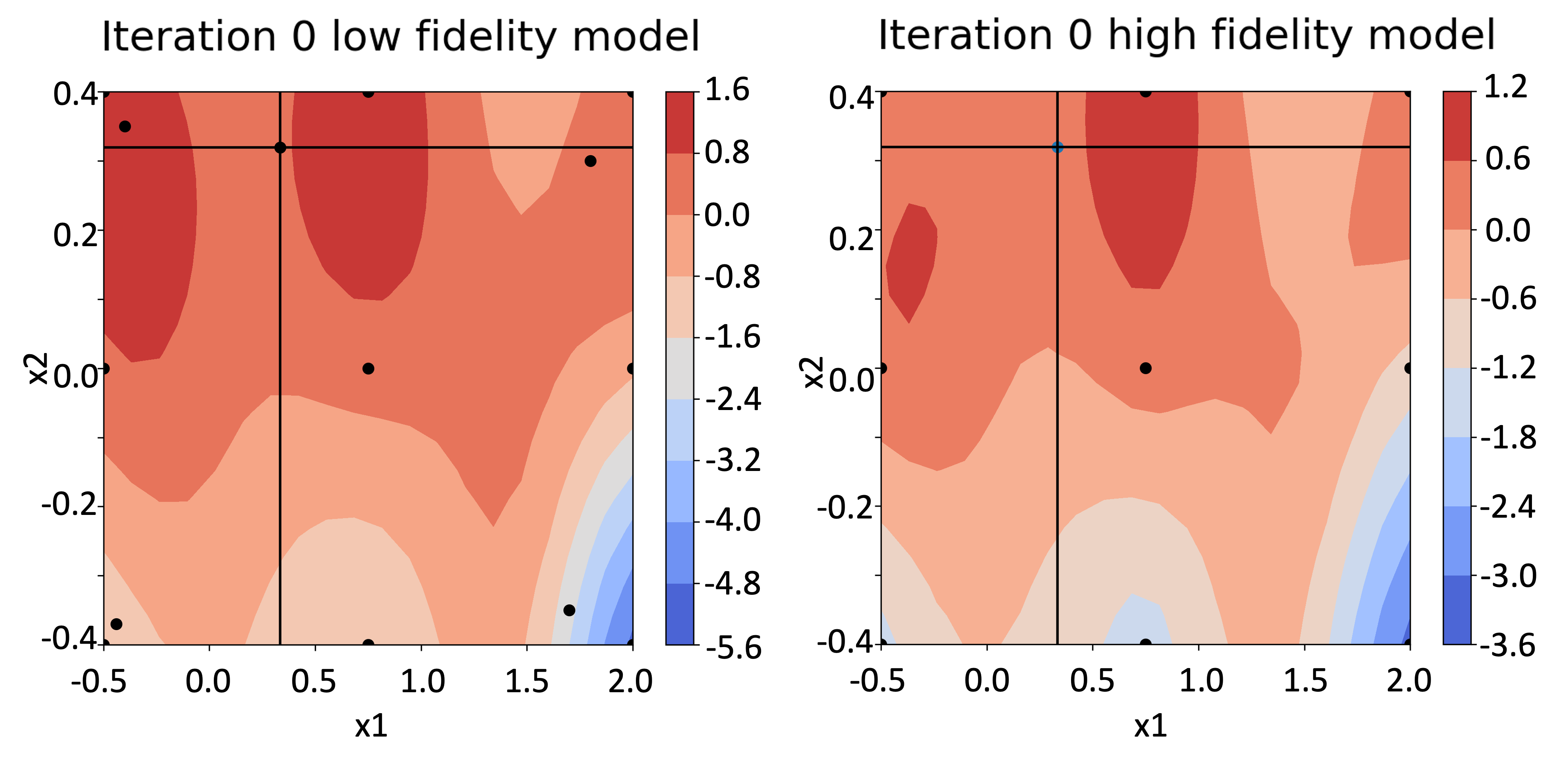}
	\caption{Mean value contour and high-fidelity sampling at iteration 0.}
\label{fig:ex1-hf-it0}
\end{figure}
\begin{figure}
  \centering
	\includegraphics[width=0.96\columnwidth]{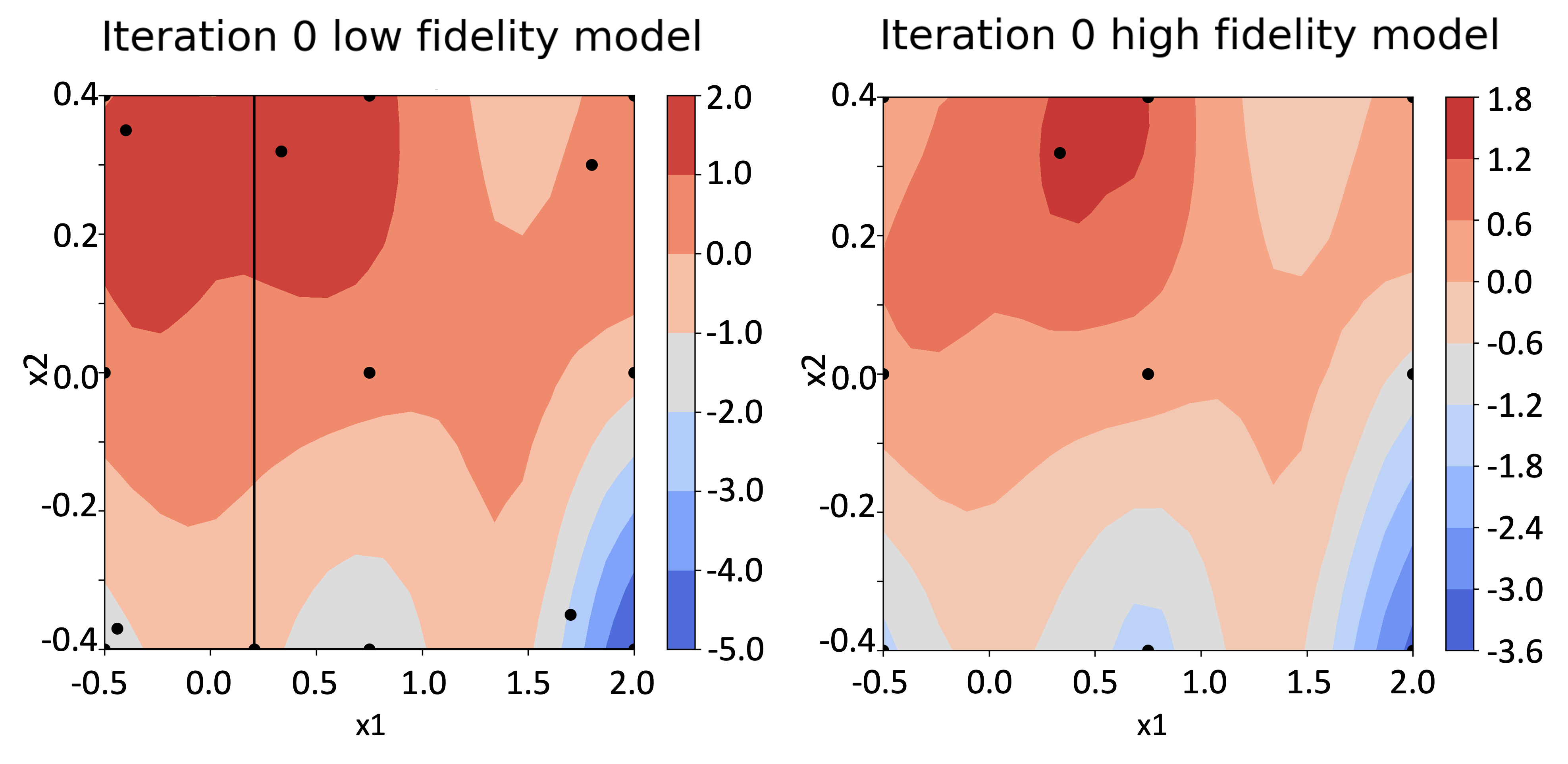}
	\caption{Mean value contour and low-fidelity sampling at iteration 0.}
\label{fig:ex1-lf-it0}
\end{figure}

The initial sampling points are nine evenly distributed grid points. The mean value contours and the sampling at $x_i^h$ for both fidelities at iteration $0$ are shown in Figure~\ref{fig:ex1-hf-it0}, where all sampled points are marked and the next sampling point is at the intersection of the lines. The sampling of $x_i^l$ at iteration $0$ is shown in Figure~\ref{fig:ex1-lf-it0}.

\begin{figure}
  \centering
	\includegraphics[width=0.7\columnwidth]{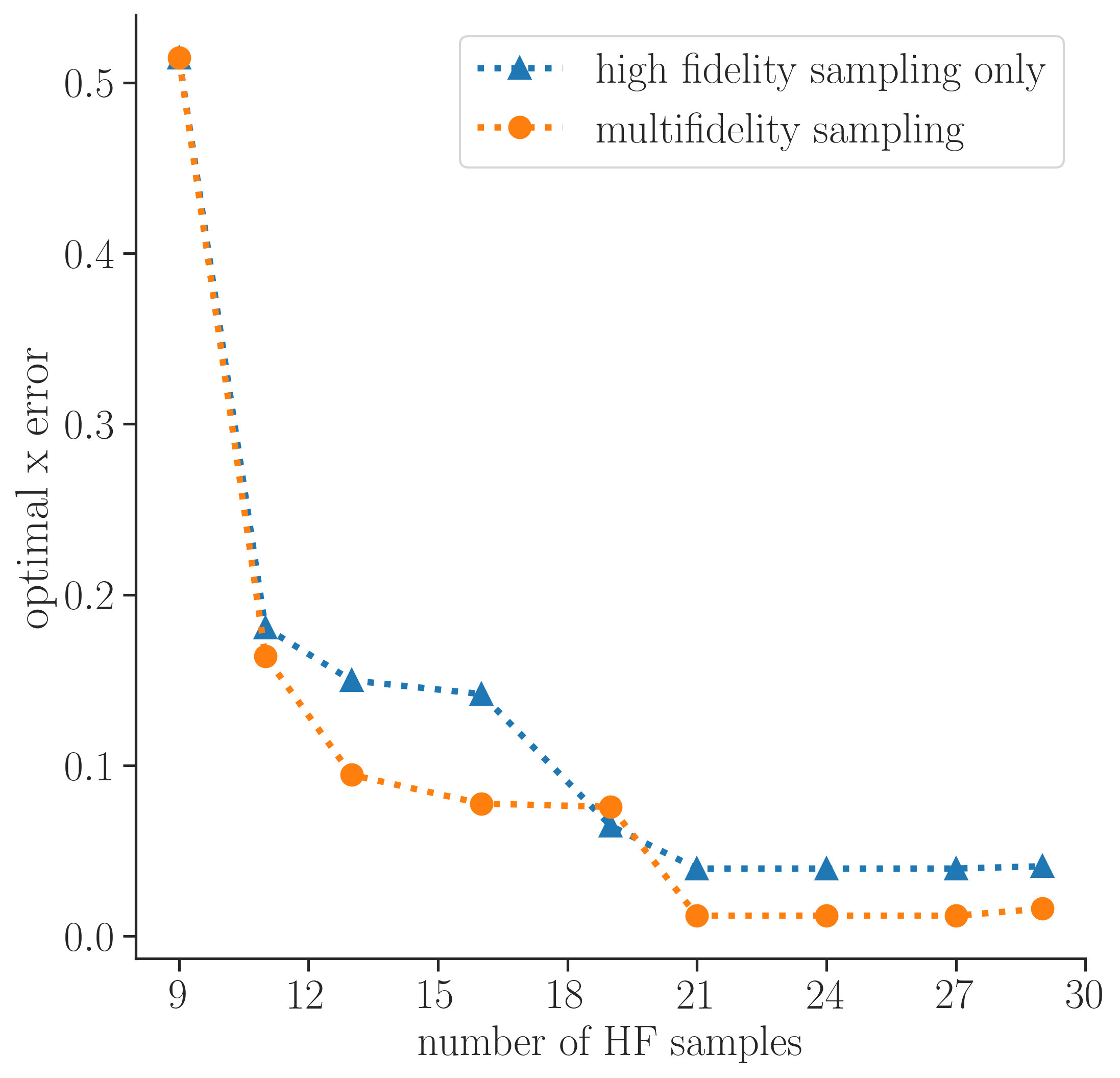}
	\includegraphics[width=0.7\columnwidth]{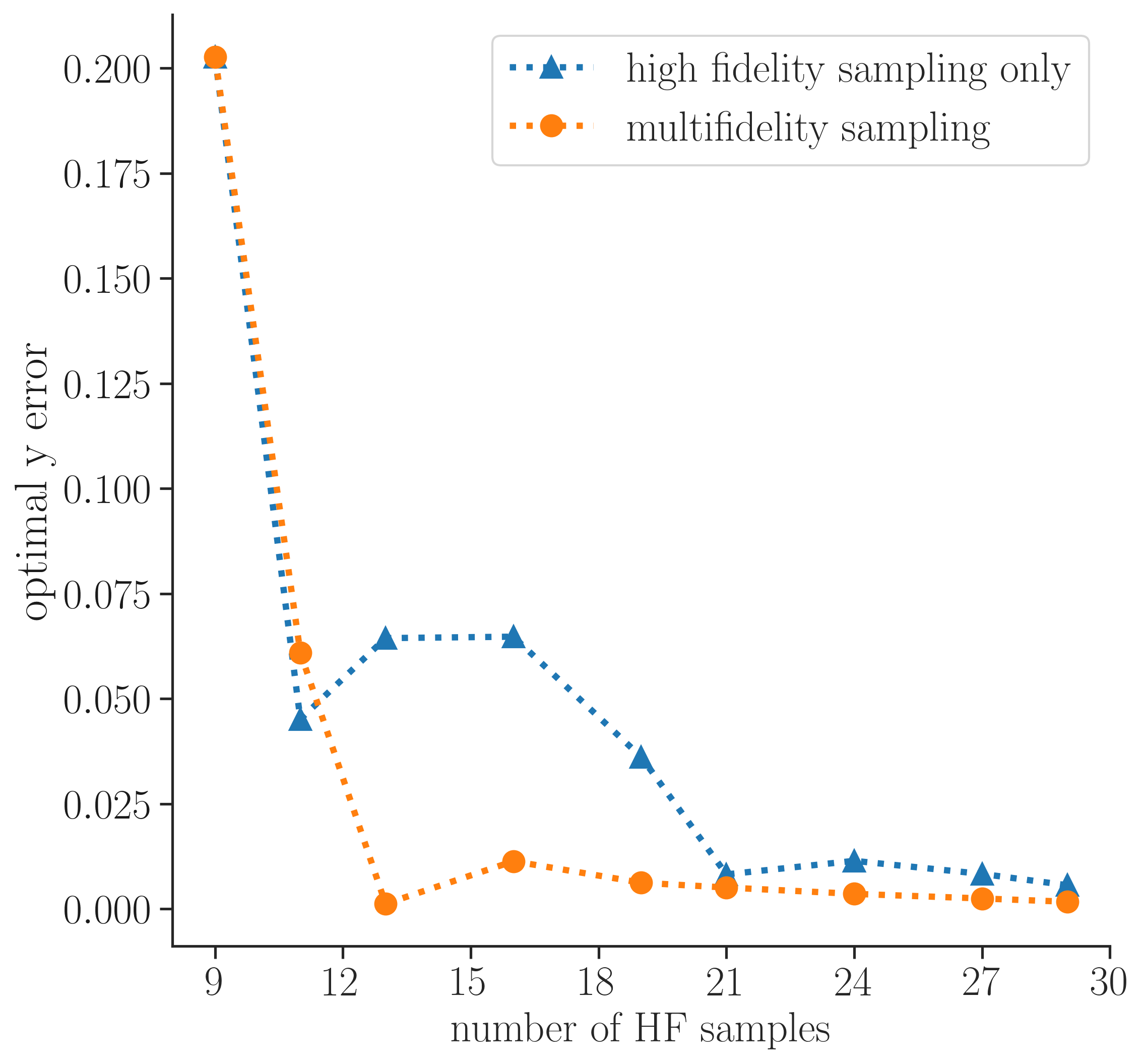}
	\caption{Our multifidelity approach produces equivalent or smaller errors in both design values (top) and output yields (bottom) as a strictly high-fidelity approach.  In particular, for a fixed number of high fidelity samples (horizontal axis), the additional information provided by the low fidelity samples does not increase the error and often decreases it.}
\label{fig:ex1-hfvsmf}
\end{figure}

The 2D example demonstrates the validity of our Bayesian optimization algorithm.
To demonstrate the potential benefit of multifidelity modeling, we compare the multifidelity results with single fidelity modeling that relies on high fidelity simulation and sampling alone, \textit{i.e.}, using Algorithm~\ref{alg:bo}. 
Figure~\ref{fig:ex1-hfvsmf} shows the error versus iterations of the two methods, where the error is measured against the ground truth of the high-fidelity model.
The optimal x error is the norm of the difference between the optimal design parameters predicted by the surrogate model and the ground truth. The optimal y error refers to the error of nuclear yield.  
It is obvious that additional low-fidelity samples and its surrogate model data improve the convergence behavior of the optimization algorithms for this problem. 
Another possible benefit is the reduction in uncertainty through more computationally inexpensive low-fidelity sampling reported in recent work~\cite{sarkar2019}. 

\subsection{8D Design Example}
In the second example, the proposed multifidelity Bayesian optimization method is applied to an eight-dimensional design problem, similarly using a surrogate trained on the HYDRA data set to enable statistical evaluation of the algorithm's performance. 
The design parameters are `\texttt{sc\_peak}', as in the 2D case, along with three dopant parameters, and four thickness parameters (1 of which is the thickness of the ice layer).
To validate the solution in eight dimensions without visualization, we exhaustively sampled $10^{10}$ data points in the eight-dimensional design space in both high fidelity and low fidelity. 

Since the design objective, the nuclear yield, has different ranges of values for the low- and high-fidelity models (toggling nuclear burn changes the yield significantly), we scale each to a range of $[0,1]$ for analysis.

\begin{figure}
  \centering
	\subfloat[Low fidelity HYDRA data histogram]{\includegraphics[width=1\columnwidth]{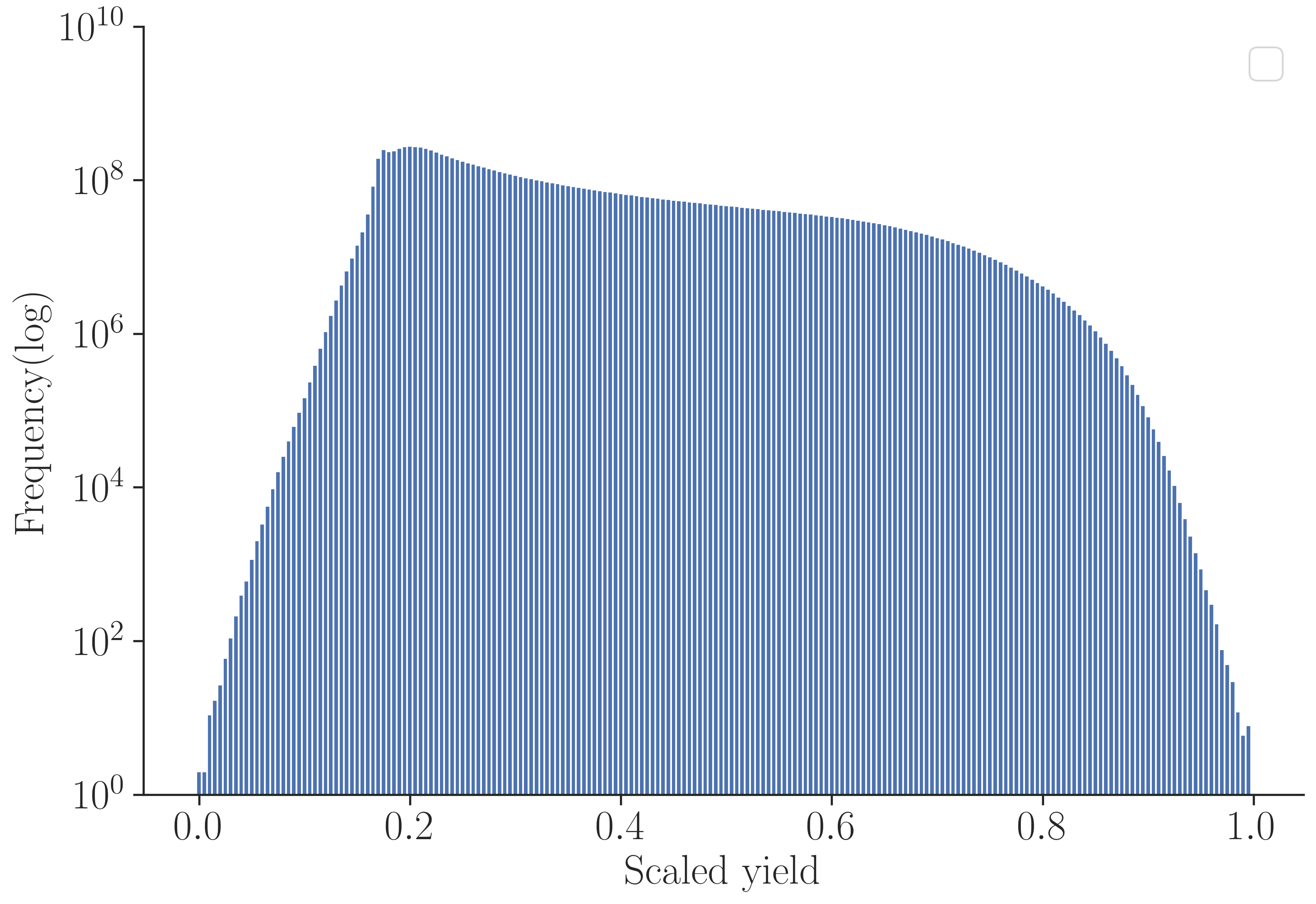}}
  \label{fig:ex2-lf}
	\subfloat[High fidelity HYDRA data histogram]{\includegraphics[width=1\columnwidth]{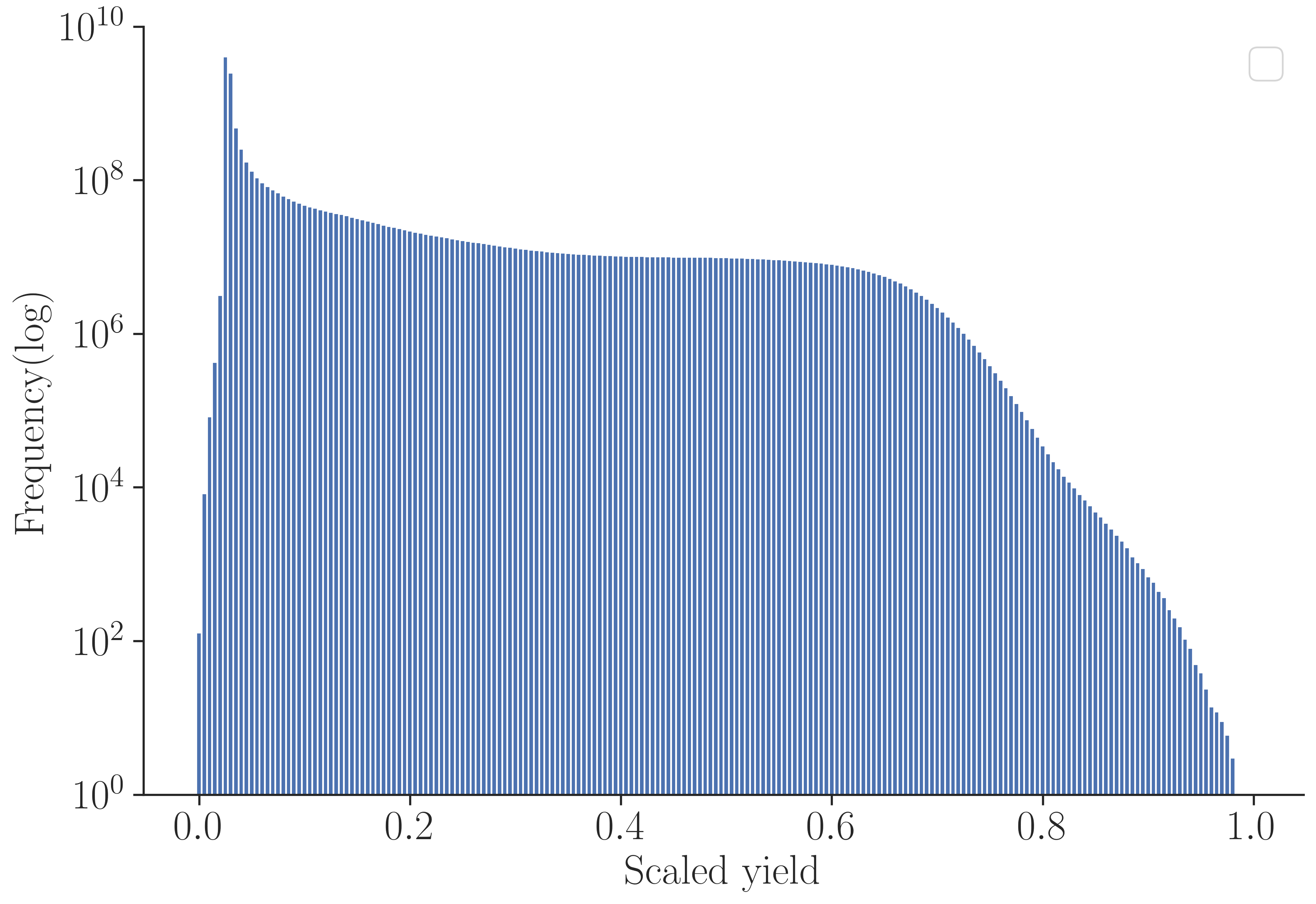}}
  \label{fig:ex2-hf}
\caption{Histogram exhaustive sampling of HYDRA surrogate functions.  The frequency counts are log-scaled.  It is evident from the histograms that the low- and high-fidelity surrogates have distinct distributions of values for the same range of parameters.}
\label{fig:ex2-histgram}
\end{figure}

Histograms of the scaled yield over $10^{10}$ samples are given in Figure~\ref{fig:ex2-histgram}. The distributions are clearly different, which supports the conclusion that the low and high-fidelity response surfaces are also different. Another quantitative measure of this appears in the population percentiles of Table~\ref{tab:ex2-percentile}. The high-fidelity burn-on simulations have a small population of very high yield, that doesn't appear in the low-fidelity population. (In other words, a small fraction likely ignite and produce significant yield.)

\begin{table}[htb]
  \begin{center}{
	  \begin{tabular}{|r|c|c|}
			\hline
			percentile & LF Yield & HF Yield \\    
			\hline
			 5th & 0.1774 & 0.0289 \\ \hline
			25th & 0.2156 & 0.0294 \\ \hline
			50th & 0.2805 & 0.0307 \\ \hline
			75th & 0.4171 & 0.0513 \\ \hline
			90th & 0.5698 & 0.4055 \\ \hline
			99th & 0.8348 & 0.7097 \\ \hline
			99.9th & 0.8808 & 0.7638 \\ \hline
		\end{tabular}
  }\end{center}
  \caption{Percentile data for exhaustive sampling of the low fidelity (LF) and high fidelity (HF) surrogate functions provides a quantitative view of the histograms from Figure~\ref{fig:ex2-histgram}. }
	\label{tab:ex2-percentile}
\end{table}

Similar to the two-dimensional case, the optimal parameter regimes for the two fidelities are quite far apart from each other. 
To demonstrate this, Table~\ref{tab:ex2-mismatch} shows the top three high-fidelity and low-fidelity values, along with the corresponding yields for the two models evaluated at those same points. The best high-fidelity points are only moderate low-fidelity performers. Even more striking, the best low-fidelity points produce very little high-fidelity yield, with the best one producing only 2\% of the best the high-fidelity point.

\begin{table}[htb]
  \begin{center}{
	  \begin{tabular}{|r|c|c|}
			\hline
			Best HF solution & LF Yield & HF Yield\\    
			\hline
			 1 & 0.5729 & 1.0000 \\ \hline
			 2 & 0.5903 & 0.9923 \\ \hline
			 3 & 0.5778 & 0.9857 \\ \hline
			 \hline
			Best LF solution & LF Yield & HF Yield\\    
			 1 & 1.0000 & 0.0214 \\ \hline
			 2 & 0.9675 & 0.2888 \\ \hline
			 3 & 0.9655 & 0.1183 \\ \hline
		\end{tabular}
  }\end{center}
  \caption{From our exhaustive search, we find that the highest yields found by the low-fidelity and high-fidelity models are not co-located in parameter space.}
	\label{tab:ex2-mismatch}
\end{table}

Just as was found in two dimensions, the optimal low-fidelity design does not correspond with the optimal high-fidelity design, motivating the need for an iterative multifidelity approach to optimization.

\begin{figure}
  \centering
	\includegraphics[width=0.95\columnwidth]{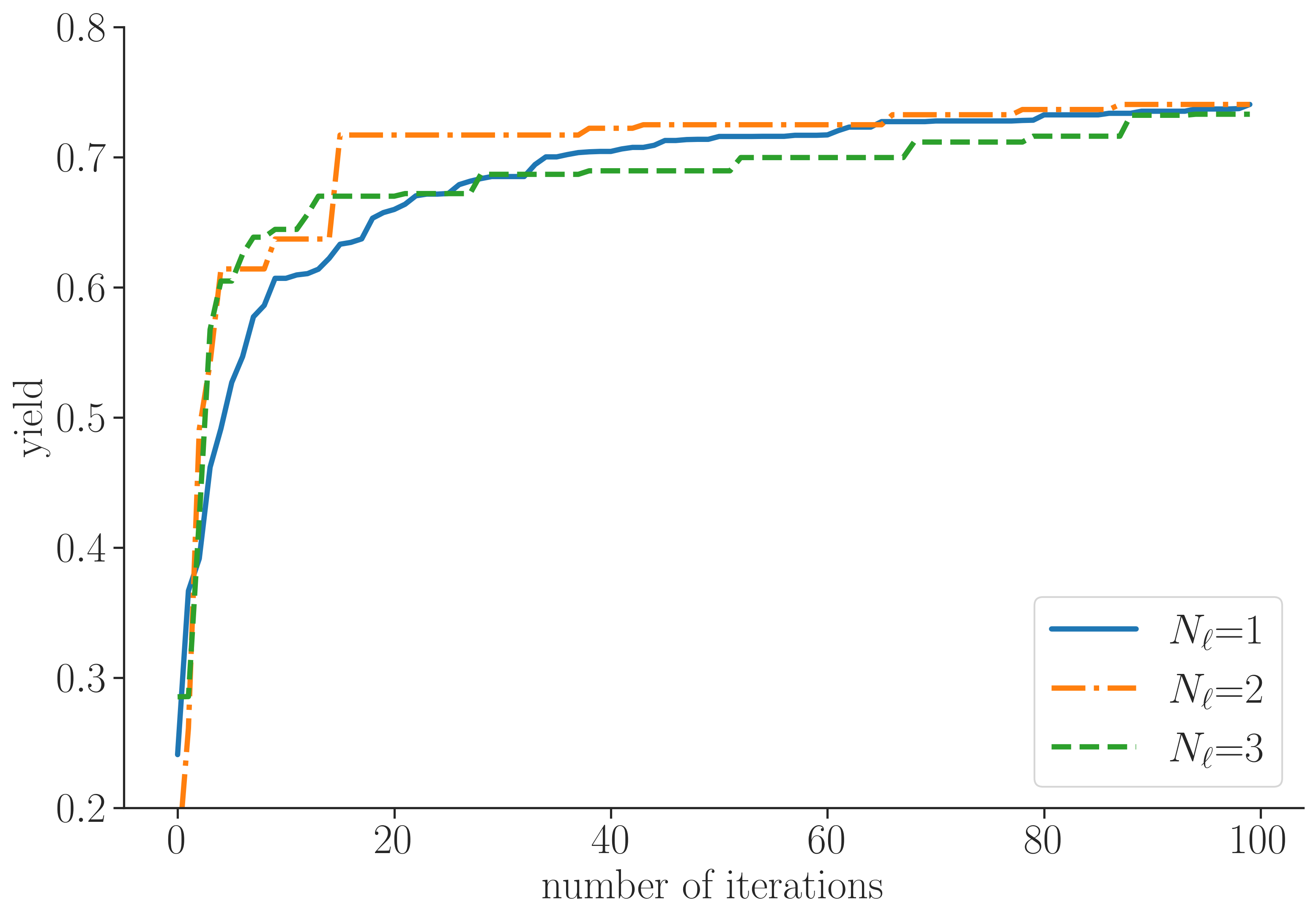}
	\caption{The yield attained over 100 iterations of our multifidelity Bayesian optimization algorithm for the 8D example is similar for three choices of $N_\ell$.  All three choices achieve a yield value in the 99th percentile of known values; using $N_2$ produces a slightly better result than $N_1$ or $N_3$, but all find very good solutions.}
	\label{fig:ex2-iteration}
\end{figure}

To test our algorithm in eight dimensions, we begin with a Latin hypercube sampling of 100 high-fidelity and 100 additional low-fidelity sample points and then run three different tests of 100 iterations varying $N_\ell$ from 1 to 3. As mentioned in Section~\ref{se:multi-bayesian}, both fidelities are evaluated at the candidate points of each iteration. The total number of high-fidelity simulations in each test is 200 (100 initial LHS + 100 iterations). However, the total number of low-fidelity simulations varies with $N_\ell$ such that $N_{\text{total}}=N_{\text{LHS}}+N_\ell N_{\text{iterations}}=100(1+N_\ell)$. To acquire statistics, we repeat each test ten times varying the initial sampling and take the average of the output yield, which is plotted in Figure~\ref{fig:ex2-iteration} against the number of iterations.

In all three cases, the output yield reaches the 99th percentile of the yield value, (corresponding to values above 0.7097 as in Table~\ref{tab:ex2-percentile}).
Meanwhile, increasing number of low fidelity samples seems to accelerate the optimization process during the early iterations. However, all three runs converge to similar values towards the end of the 100 iterations, with $N_\ell=2$ displaying the best optimal yield. The multifidelity Bayesian optimization algorithm is able to effectively find a near optimal solution in eight dimensions in a few hundred simulations.

\begin{figure}
  \centering
	\includegraphics[width=0.9\columnwidth]{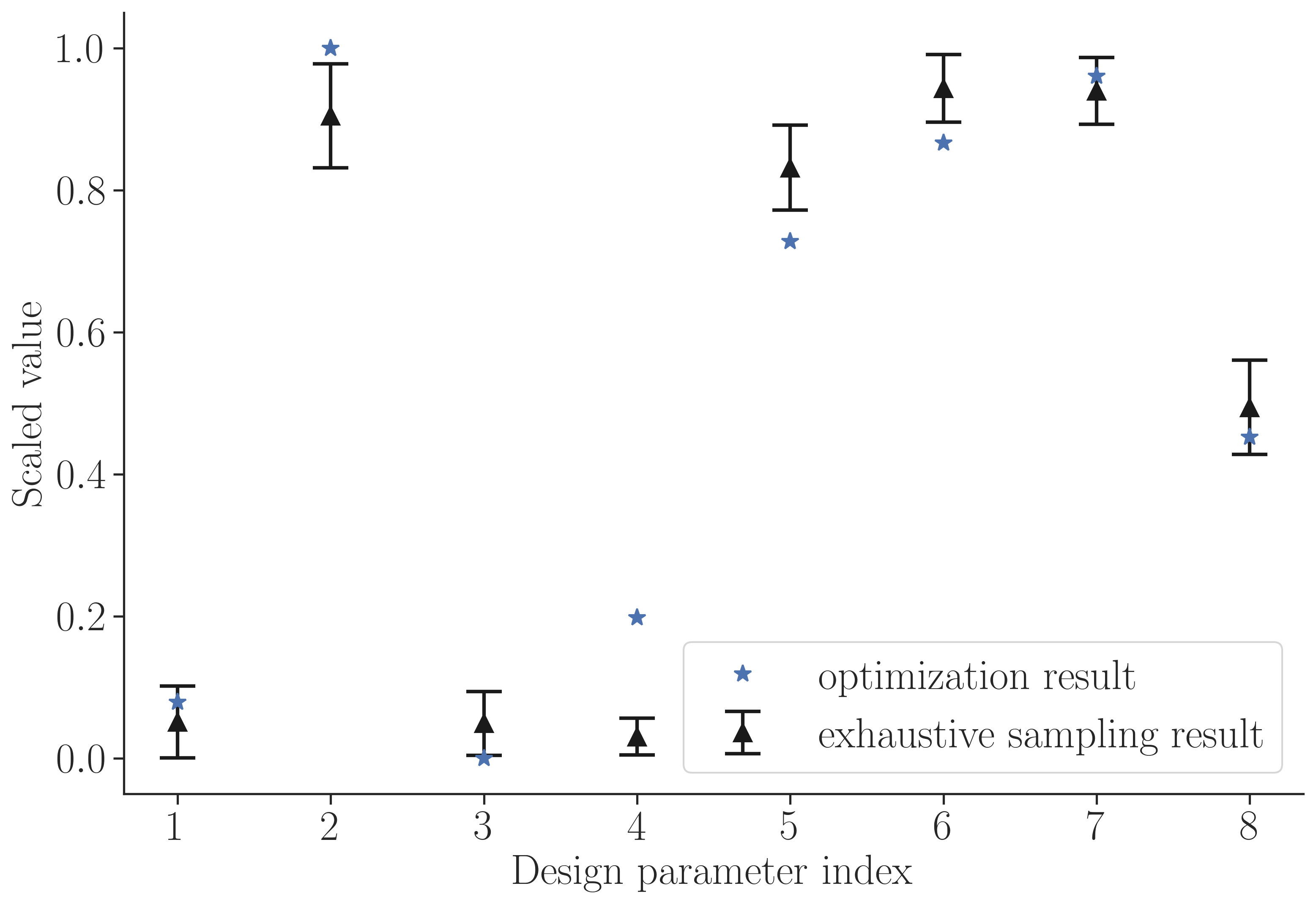}
	\caption{Comparing the design selected by our multifidelity Bayesian optimization algorithm (blue stars) to a distribution of 1000 designs with highest yield from exhaustive sampling (black triangles with error bars) suggests that the design is close to optimal.}
	\label{fig:ex2-xopt}
\end{figure}

Since each run converges on different optima, there may be sensitivity to initial sampling. To investigate this, we rerun the $N_\ell=1$ algorithm for ten different initial LHS designs and look at the final location of each run in parameter space (in other words, to see whether random starts end up near each other or converge to disparate optima). The average (coordinate wise) of these ten runs is compared with the mean and standard deviation of the coordinates of the top 1000 simulations from the exhaustive 10 billion search in Figure~\ref{fig:ex2-xopt}.

We first note that there seems to exist an optimal region for each of the eight design parameters, as can be seen by the relatively small standard deviation among the exhuastive sampling result. Furthermore, with $200$ high-fidelity and $300$ low-fidelity samples chosen via the multifidelity Bayesian optimization, our algorithm produces an optimal design that is very near the optimal region given by exhaustive sampling, in almost all coordinates within two standard deviations of the optimal region.

In both two and eight dimensions, our iterative multifidelity algorithm can effectively navigate a complicated ICF design landscape, while using low-fidelity models to accelerate the search for good designs.

%% file: Sections_journal/Conclusion.tex
\section{Conclusions}
\label{sec:con}
In this work, we address the problem of multifidelity design for ICF.
We show that even in a simple case, the design landscape changes with fidelity. As such, a two-step optimization approach that generates a local design region from low-fidelity models would guide the high-fidelity simulation to less optimal design parameters. In other words: a two-stage optimization approach might not be effective for ICF design. As the physics changes between fidelities, so too must the design landscape (and therefore the optimal design).  And although this discrepancy also holds true in the higher dimensional design space, it should not be too surprising: higher simulation fidelity has been pursed in ICF because of this exact problem. If simple models could accurately predict the best design, there would be little need for additional sophistication. One could merely optimize a one-dimensional simulation and expect it to perform as expected in an experiment.

Furthermore, our result questions the efficacy of experimental campaigns that propose to find optimal designs with THD capsules and then test them with DT. The ``best" THD design may be misleading, with parameters far away from the ``best" DT design.

However, we present in this paper an iterative multifidelity Bayesian optimization method based on Gaussian Process surrogate models to address this issue.
We select well-established kernel, acquisition functions, and optimization algorithms that allow for the dynamic allocation of simulation resources. This not only helps guard against the possibility of being mislead by low-fidelity designs, but also allows for efficient use of low and high-fidelity models. We preferentially explore with low-fidelity simulations and exploit with high-fidelity simulations. And although developed and tested with simulations, our algorithm could be applied to experimental or mixed simulation-experimental campaigns.

Our algorithm shows a statistically significant improvement over single-fidelity search, making it a promising option for efficient design.